\documentclass[twocolumn,aps,graphicx,prb,superscriptaddress]{revtex4-1}

\usepackage{amsmath}
\usepackage{amsfonts}
\usepackage{graphicx}  
\usepackage{float}  
\usepackage{subfigure} 
\usepackage{epstopdf}
\usepackage{epsfig}
\usepackage{color}
\usepackage[colorlinks,citecolor=blue]{hyperref}
\usepackage{multirow}
\usepackage{natbib}
\usepackage{tabularx}
\usepackage{color,ulem}

\begin{document}

\title{Two-dimensional Kagome-in-Honeycomb materials (MN$_4$)$_3$C$_{32}$ (M=Pt or Mn)}

\author{Jingping Dong}
\affiliation{The Center for Advanced Quantum Studies and Department of Physics, Beijing Normal University, Beijing 100875, China}
\affiliation{Key Laboratory of Multiscale Spin Physics (Ministry of Education), Beijing Normal University, Beijing 100875, China}

\author{Miao Gao}
\affiliation{Department of Physics, School of Physical Science and Technology, Ningbo University, Zhejiang 315211, China}

\author{Xun-Wang Yan}\email{yanxunwang@163.com}
\affiliation{College of Physics and Engineering, Qufu Normal University, Shandong 273165, China}

\author{Fengjie Ma}\email{fengjie.ma@bnu.edu.cn}
\affiliation{The Center for Advanced Quantum Studies and Department of Physics, Beijing Normal University, Beijing 100875, China}
\affiliation{Key Laboratory of Multiscale Spin Physics (Ministry of Education), Beijing Normal University, Beijing 100875, China}

\author{Zhong-Yi Lu}
\affiliation{Department of Physics, Renmin University of China, Beijing 100872, China}
\affiliation{Key Laboratory of Quantum State Construction and Manipulation (Ministry of Education), Renmin University of China, Beijing 100872, China}

\date{\today}

\begin{abstract}

We propose two novel two-dimensional (2D) topological materials, (PtN$_4$)$_3$C$_{32}$ and (MnN$_4$)$_3$C$_{32}$, with a special geometry that we named as kagome-in-honeycomb (KIH) lattice structure, to illustrate the coexistence of the paradigmatic states of kagome physics, Dirac fermions and flat bands, that are difficult to be simultaneously observed in three-dimensional realistic systems. In such system, MN$_4$(M=Pt or Mn) moieties are embedded in honeycomb graphene sheet according to kagome lattice structure, thereby resulting in a KIH lattice. Using the first-principles calculations, we have systemically studied the structural, electronic, and topological properties of these two materials. In the absence of spin-orbit coupling (SOC), they both exhibit the coexistence of Dirac/quadratic-crossing cone and flat band near the Fermi level. When SOC is included, a sizable topological gap is opened at the Dirac/quadratic-crossing nodal point. For nonmagnetic (PtN$_4$)$_3$C$_{32}$, the system is converted into a $\mathbb{Z}_2$ topological quantum spin Hall insulator defined on a curved Fermi level, while for ferromagnetic (MnN$_4$)$_3$C$_{32}$, the material is changed from a half-semi-metal to a quantum anomalous Hall insulator with nonzero Chern number and nontrivial chiral edge states. Our findings not only predict a new family of 2D quantum materials, but also provide an experimentally feasible platform to explore the emergent kagome physics, topological quantum Hall physics, strongly correlated phenomena, and theirs fascinating applications.

\end{abstract}

\maketitle

\section{INTRODUCTION}

Two-dimensional (2D) materials have attracted widespread attention in recent years in the fields of materials science and condensed matter physics, owing to their novel physical properties and great importance for fundamental research and potential applications \cite{doi:10.1126/science.1102896,doi:10.1126/science.1165429,doi:10.1126/science.1165429,GUPTA201544,10.1093/nsr/nwu080,https://doi.org/10.1002/wcms.1313,Jing2020,Galeotti2020,Liu2020,PhysRevB.106.064407}. Nowadays, 2D materials have become a large family with various categories, including elemental 2D materials \cite{Mannix2017,glavin2020emerging,https://doi.org/10.1002/adfm.202107280}, transition metal dichalcogenides \cite{Wang2012,Xu2013,Lv2015,Manzeli2017}, MXenes \cite{https://doi.org/10.1002/adma.201102306,https://doi.org/10.1002/adma.201304138,Bhat2021}, metal-organic frameworks \cite{james2003metal,zhou2012introduction,furukawa2013chemistry,Dong2018}, transition metal carbonitride \cite{doi:10.1126/science.aba7977,PhysRevB.103.125407,PhysRevB.103.155411,PhysRevMaterials.6.074202,Feng2022,D3CP00407D}, etc. Different degrees of freedom in 2D systems, including spin, valley, orbit, topology, and layer, may become entangled with each other, giving rise to intriguing new physics and applications \cite{Du2021,Schaibley2016}. The search for new types of 2D materials can provide exciting opportunities to explore new physics or applications and is therefore an important research topic \cite{Mueller2018,Lin2023,https://doi.org/10.1002/smsc.202100033}. 

Most recently, a new class of materials called kagome materials has opened the door to investigate the emergent physical properties resulting from the quantum interactions between geometry, topology, spin, and correlation \cite{PhysRevLett.108.045305,PhysRevLett.120.026801,Ye2018,Xue2019,Guguchia2020,Yin2022,Wang2023,Guguchia2023,Yin2019,Ghimire2020,Hu2023,PhysRevB.85.195320,Xu2015,Pan2017,Yang2009,PhysRevLett.114.245504,Wakefield2023,You2022}. The coexistence of topological flat bands and Dirac fermions renders the kagome materials as one of the most interesting platform for studying strong correlated phenomena and topological non-trivial properties. In this system,  the Dirac fermions interconnect with topology, while flat bands with quenched kinetic energy facilitate strongly correlated phenomena. Their combination in one system creates a great opportunity for the realization and discovery of a series of advanced materials with exotic properties, such as the fractional quantum Hall effect \cite{PhysRevLett.106.236802, PhysRevLett.48.1559, PhysRevLett.107.126803, PhysRevLett.50.1395} and high-Tc superconductivity \cite{PhysRevLett.126.247001,Chen2021, Zhao2021,Li2022}. However, to our best knowledge, the paradigmatic states of the idealized kagome lattice, Dirac fermions and flat bands, have not been simultaneously observed in realistic materials. The Dirac fermions are usually observable due to the topological protections \cite{PhysRevB.85.195320,Xu2015,Pan2017}, while flat bands are parametric rather than topological, making them difficult to observe in practice \cite{Yang2009,PhysRevLett.114.245504,Wakefield2023}. In the reported kagome materials, the complex three-dimensional and multi-order electron hoppings result in the disappearance of ideal kagome features \cite{You2022}.

\begin{figure}
\centering
\includegraphics[width=1.0\linewidth]{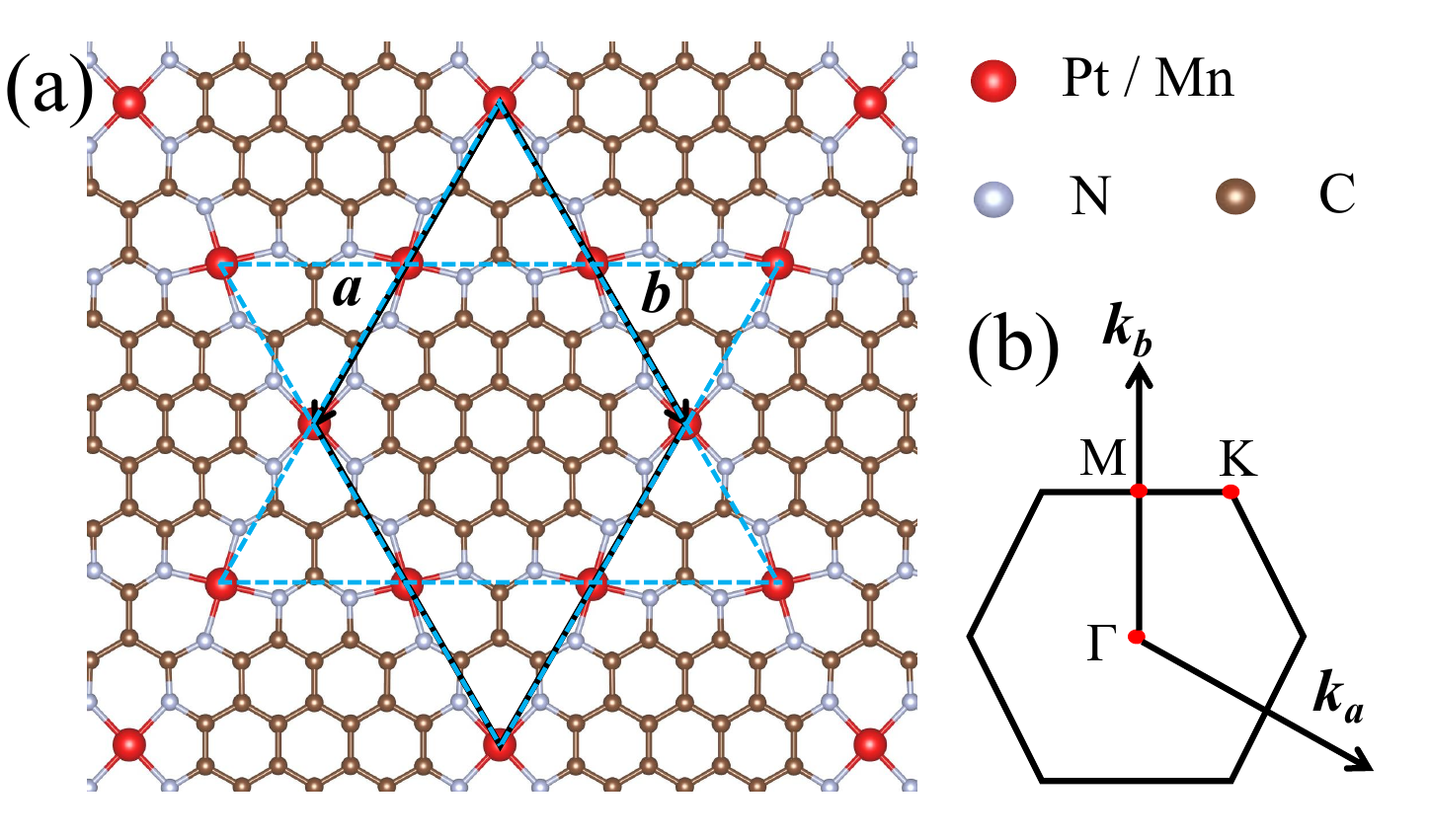}
\caption{\label{crystal}  (a) Atomic structure of monolayer KIH materials (MN$_4$)$_3$C$_{32}$ (M= Pt or Mn), in which the red, gray, and brown balls represent M (Pt or Mn), N, and C atoms, respectively. The black solid lines give the primitive cell, while the light blue dashed lines illustrate the kagome lattice formed by the MN$_4$ moieties.  (b) The corresponding Brillouin zone of (MN$_4$)$_3$C$_{32}$ (M= Pt or Mn) with high-symmetry k-points labeled.}
\end{figure}

Here, we propose an alternative way to construct kagome physics in original non-kagome materials (such as graphene) by embedding MN$_4$ (M = metal) moieties in a form of kagome lattice way, coined as a kagome-in-honeycomb (KIH) lattice, as shown in the Fig. \ref{crystal}(a).  In fact, embedding MN$_4$ moieties into graphene has become practical in experiments of single-atom catalysts synthesis \cite{deng2015, Fei2018, He2019, Zhao2019, Dou2020}. Until now, more than 20 kinds of MN$_4$ moieties have been successfully embedded into graphene, as identified by systematic X-ray absorption fine structure analyses and direct transmission electron microscopy imaging \cite{Fei2018}.

In the prototypical KIH materials (MN$_4$)$_3$C$_{32}$ (M=Pt or Mn), both of a Dirac/quadratic-crossing cone and a flat band, signatures of topology and correlation, are found in the vicinity of Fermi level.  The superposition of the honeycomb and the kagome lattice would show the characteristic band structures superposed by those of constituting sub-lattices separately, and hence, much rich electronic and topological properties are expected in the new type of 2D materials. By the first-principles calculations, we predict that (PtN$_4$)$_3$C$_{32}$ exhibits a significant quantum spin Hall effect and holding great promise for experimental realization. In the absence of SOC, it is a perfect gapless Dirac semimetal, with the Dirac points locate at particularly symmetric K and K$'$ points on the Brillouin zone boundary. Once SOC is included, a  local gap is opened at the Dirac point, and the system is converted into a $\mathbb{Z}_2$ topological insulator defined on a curved Fermi level. In addition, we find that (MnN$_4$)$_3$C$_{32}$ is a topological half-semi-metal, with the coexistence of fully spin-polarization and quadratic dispersion nodal states. Distinguished from the nonmagnetic Dirac material (PtN$_4$)$_3$C$_{32}$, the four-fold degeneracy neck crossing-point traces out two-fold degeneracy lines emerged in the single-spin channel. When considering SOC, (MnN$_4$)$_3$C$_{32}$ becomes a quantum anomalous Hall insulator with its energy gap of E$_g$ = 13 meV, which is characterized by the nonzero Chern number (C = 1) and chiral edge states.

\section{METHODS}

In our calculations, plane-wave basis based methods were used. We employed the projector-augmented plane-wave (PAW) method \cite{2001All,PhysRevB.59.1758,PhysRevB.50.17953} as implemented in the Vienna Ab-initio Simulation Package (VASP) \cite{PhysRevB.54.11169}. The PAWs were generated from the Perdew-Burke-Ernzerhof (PBE) functional \cite{perdew1996generalized}. For the (PtN$_4$)$_3$C$_{32}$ calculations, we adopted the generalized gradient approximation (GGA) of PBE formula for the exchange-correlation potentials in the electronic structure simulations, while for the (MnN$_4$)$_3$C$_{32}$ calculations involving partially filled 3$d$ orbitals, a corrective Hubbard-like $U$ term was introduced to treat the strong on-site Coulomb interaction of localized electrons of the transition metal Mn atoms \cite{PhysRevB.71.035105}. The effective value of U was set to 4.0 eV \cite{PhysRevB.73.195107, PhysRevB.84.045115}.  The energy cutoff energy of the plane waves is set to 500 eV with a precision energy of 10$^{-7}$ eV. A slab geometry was applied, where we added a vacuum space of $\sim$ 16 $\AA$ in the $z$ direction to eliminate the periodic effect. The mesh of \mbox{k-points} grid used for sampling the Brillouin zone was \mbox{6$\times$6$\times$1}. During the simulations, all structural geometries were fully optimized to achieve the minimum energy. The surface states were studied using tight-binding methods by the combination of Wannier90 \cite{mostofi2008wannier90} and WannierTools \cite{WU2017} software packages.

\section{RESULTS AND DISCUSSION}

\subsection{(PtN$_4$)$_3$C$_{32}$}

\begin{figure}
\centering
\includegraphics[width=1.0\linewidth]{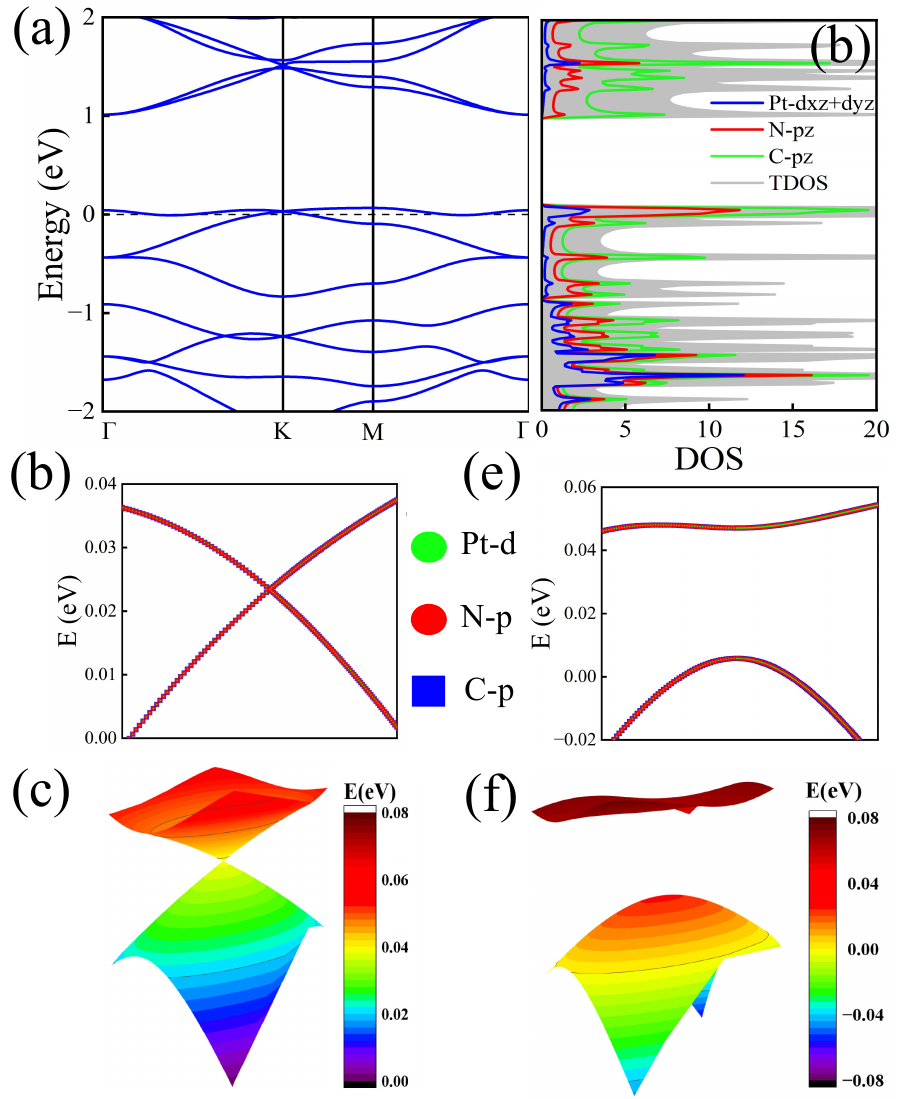}
\caption{\label{bands} (a) Band structure of monolayer (PtN$_4$)$_3$C$_{32}$ without SOC. The black dashed line represents the Fermi level. The Fermi level is set to zero. (b) and (e) give the zoomed-in view of the areas around the Dirac point without SOC and with SOC, respectively. The Dirac node is found to mainly consist of the N-$p$, C-$p$, and Pt-$d$ states. (c) and (f) show the corresponding three-dimensional energy dispersion of bands without SOC and with SOC, respectively. (d) The density of states (DOS) of (PtN$_4$)$_3$C$_{32}$. }
\end{figure}

The crystal structure of monolayer (PtN$_4$)$_3$C$_{32}$ belongs to $P6/mmm$ (No. $191$) space group, which is affiliated to $D_{6h}$ point group symmetry, as shown in Fig. \ref{crystal}(a). There are three Pt,  twelve N, and thirty-two C atoms in the primitive cell (indicated by the solid black rhombus). The Pt atoms together with their four neighboring coordinating N atoms form PtN$_4$ moieties, which are connected by C atoms to build the KIH lattice. After structural relaxation, all three kinds of atoms in monolayer (PtN$_4$)$_3$C$_{32}$ are coplanar, with the optimized lattice parameters of $a$ = $b$ = 12.37 $\AA$. Figure \ref{crystal}(b) illustrates the corresponding Brillouin zone of (PtN$_4$)$_3$C$_{32}$ with high symmetry points labeled, which is similar to that of a perfect honeycomb structure.

To study the electronic properties of monolayer (PtN$_4$)$_3$C$_{32}$, we investigate its band structure and the corresponding total and partial density of states (TDOS and PDOS). Figure \ref{bands}(a) shows the calculated electronic band structure. Similar to graphene, there are two bands linearly crossing at the high-symmetry K/K$'$ point around the Fermi level, indicating that monolayer (PtN$_4$)$_3$C$_{32}$ is a Dirac semimetal in the absence of SOC. Figures \ref{bands}(b) and \ref{bands}(c) show the zoomed-in view of the area and the three-dimensional energy dispersion of highest valence and lowest conduction bands around the Dirac points. The two bulk bands touch each other only at K/K$'$ point, forming six Dirac cones at the corners of the first Brillouin zone. Moreover, a flat band is emerged in the band structure of monolayer (PtN$_4$)$_3$C$_{32}$, which is a typical characteristic of kagome-like lattice. It should be emphasized that, unlike general kagome materials, such flat band appears exactly at the Fermi level in monolayer (PtN$_4$)$_3$C$_{32}$. The coexistence of Dirac cone and flat band at the Fermi level is very convenient for exotic quantum applications. Due to the emergence of flat band, there is a sharp peak around the Fermi level in the TDOS and PDOS, as shown in Fig. \ref{bands}(d).  These states are mainly composed of N-$p$, C-$p$, and Pt-$d$ orbitals, and more specifically, the C-$p_z$ and N-$p_z$ form a $\pi$ network  in the vicinity of the Fermi level together with the Pt $d_{xz}$ and $d_{yz}$ orbitals.

\begin{figure}
\centering
\includegraphics[width=1.0\linewidth]{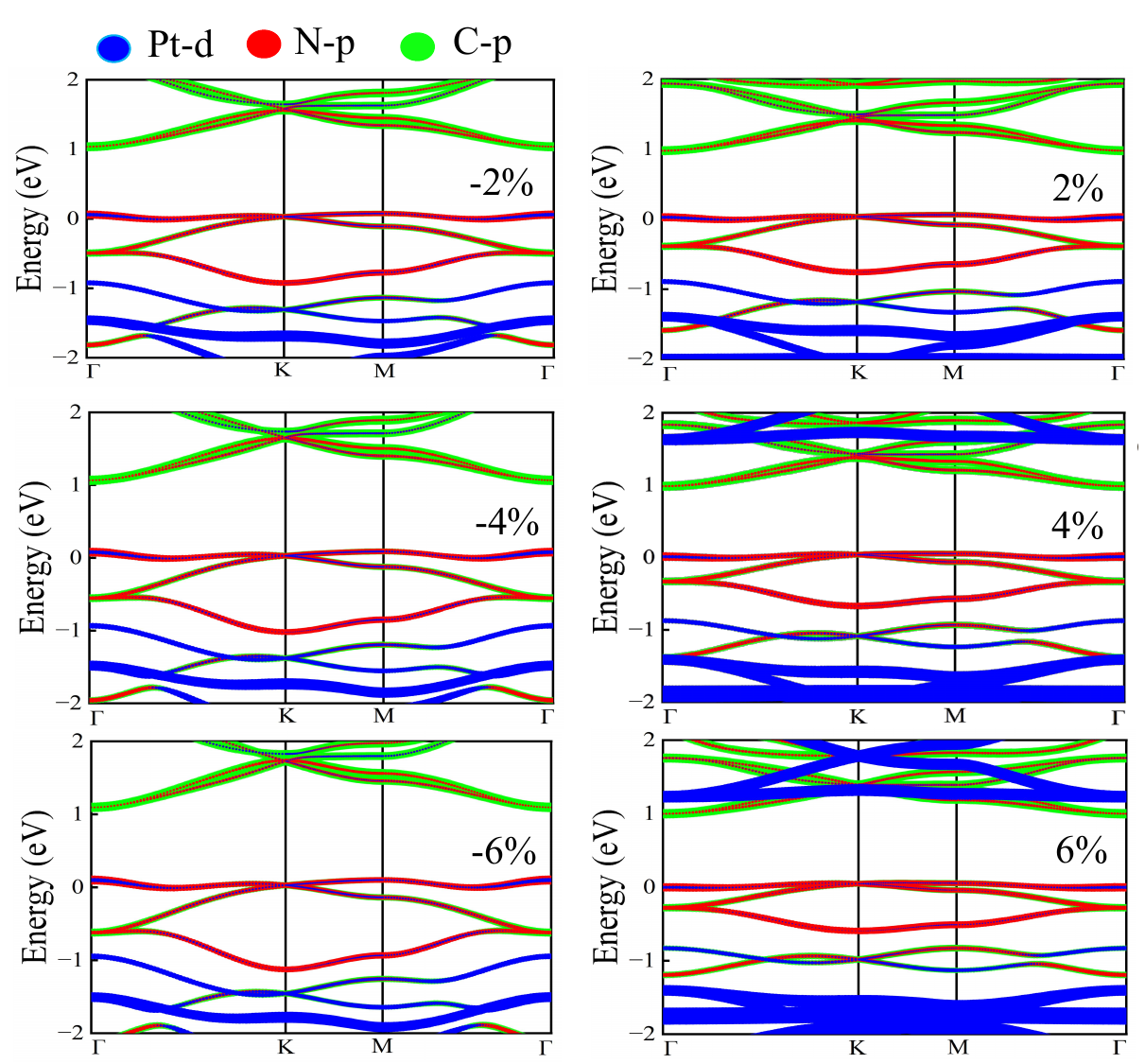}
\caption{\label{strain} Orbital-resolved band structures of the (PtN$_4$)$_3$C$_{32}$ monolayer under mild compressional or tensile strains (-2\%, -4\%, -6\% and 2\%, 4\%, 6\%), under which the evident maintenance of Dirac cones persists. The blue, red and green dots represent the contributions from the Pt-$d$, N-$p$, C-$p$ atomic orbitals, respectively. }
\end{figure}

To further evaluate the robustness of the Dirac cone and flat band against the structural deformation, biaxial tensile/compressive strains up to $\pm 6\%$ are exerted on the monolayer (PtN$_4$)$_3$C$_{32}$, as shown in Fig. \ref{strain}. Under a wide range of strains, the flat band always exists, as well as the degenerate linearly-crossing bands at the high-symmetry K/K$'$ point near the Fermi level. The in-plane strains also do not affect the main contributions of the N-$p$,  C-$p$ and Pt-$d$ orbitals in the Dirac cones. This insensitivity to the in-plane strains would facilitate the potential applications of monolayer (PtN$_4$)$_3$C$_{32}$ in more complicated mechanical situations.

Once SOC is included, a local gap is opened at the Dirac point, as shown in Fig. \ref{bands}(e). Since Pt belongs to a heavy transition-metal element, a large splitting energy gap of $\sim$40\ meV is observed, in comparison with that of graphene on the order of $\mu$eV \cite{PhysRevB.75.041401}. The splitting gap is experimentally visible and therefore is capable of hosting novel quantum spin Hall states. Figure \ref{bands}(f) shows the three-dimensional energy dispersion of highest valence and lowest conduction bands of monolayer (PtN$_4$)$_3$C$_{32}$ with SOC. The quadratic characterization with an open energy gap is very pronounced compared to the case without SOC, resulting in a more flat band.

\begin{figure}
\centering
\includegraphics[width=1.0\linewidth]{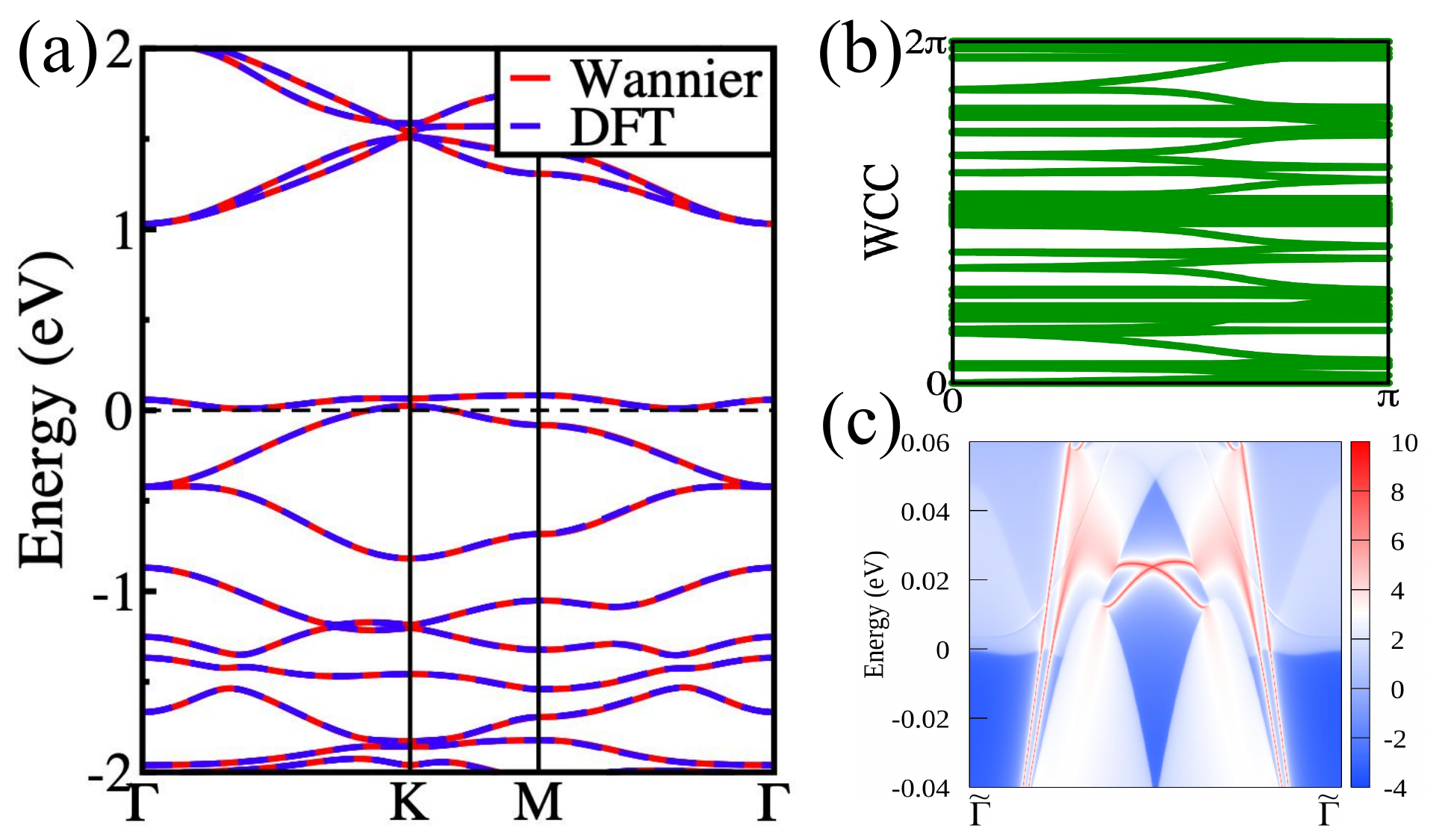}
\caption{\label{wannier} (a) The comparison of the band structure of (PtN$_4$)$_3$C$_{32}$ from the Wannier tight-binding Hamiltonian (red solid lines) and the DFT calculations (blue dashed lines).  SOC is included, and the Fermi level is set to zero.  (b) The evolution of Wannier charge center (WCC) curves in half Brillouin zone to calculate the $\mathbb{Z}_2$ invariant. (c) The helical edge and bulk band structure of (PtN$_4$)$_3$C$_{32}$ projected along the (110) direction.}
\end{figure}

From the DFT band structure with SOC, we extract a tight-bind Hamiltonian of monolayer (PtN$_4$)$_3$C$_{32}$ in the basis of maximally localized Wannier functions through the Wannier90 program \cite{RevModPhys.84.1419, mostofi2008wannier90}. Topological properties of (PtN$_4$)$_3$C$_{32}$ can then be studied with WannierTools software package based on the Green's function method \cite{sancho1984quick, WU2017}.  As shown in Fig. \ref{wannier}(a), the band structure constructed by the means of maximally localized Wannier functions agrees very well with the DFT results, during which Pt-$d_{xz}$, Pt-$d_{yz}$, N-$p_z$, and C-$p_z$ orbitals are adopted as the projections. To further confirm the topological non-triviality of monolayer (PtN$_4$)$_3$C$_{32}$, we have calculated the topological $\mathbb{Z}_2$ invariant. The method of Wannier charge center evolution in half Brillouin zone is adopted \cite{PhysRevB.84.075119, PhysRevB.83.035108}. As shown in Fig. \ref{wannier}(b), it is clear from the figure that the Wannier charge center evolution curves cut an arbitrary horizontal reference line odd times, indicating a value of $\mathbb{Z}_2 =1$. Thus the gapped (PtN$_4$)$_3$C$_{32}$ with SOC is a 2D $\mathbb{Z}_2$ topological insulator defined in a curved Fermi level.

We further investigate the helical edge states of monolayer (PtN$_4$)$_3$C$_{32}$ projected along the (110) direction, whose band dispersions are given in Fig. \ref{wannier}(c). Topological nontrivial edge states are visible in the Brillouin zone along the projected high-symmetry line $\widetilde{\Gamma}$--$\widetilde{M}$--$\widetilde{\Gamma}$. There are two edge states separately emerging from the bulk conduction/valence bands of one projected Dirac node, going across the bulk band gap, and merging into the valence/conduction continuum of the other projected Dirac node within the Brillouin zone. They cross each other at the $\widetilde{M}$ point, as protected by the time-reversal symmetry. The spin-momenta of these states are locked and protected from backscattering. These features are the main characteristics of the quantum spin Hall effect, which make monolayer (PtN$_4$)$_3$C$_{32}$ promising for low-power quantum electronic and spintronic applications in future \cite{RevModPhys.82.3045,RevModPhys.83.1057}.

\subsection{(MnN$_4$)$_3$C$_{32}$}

\begin{figure}
\centering
\includegraphics[width=1.0\linewidth]{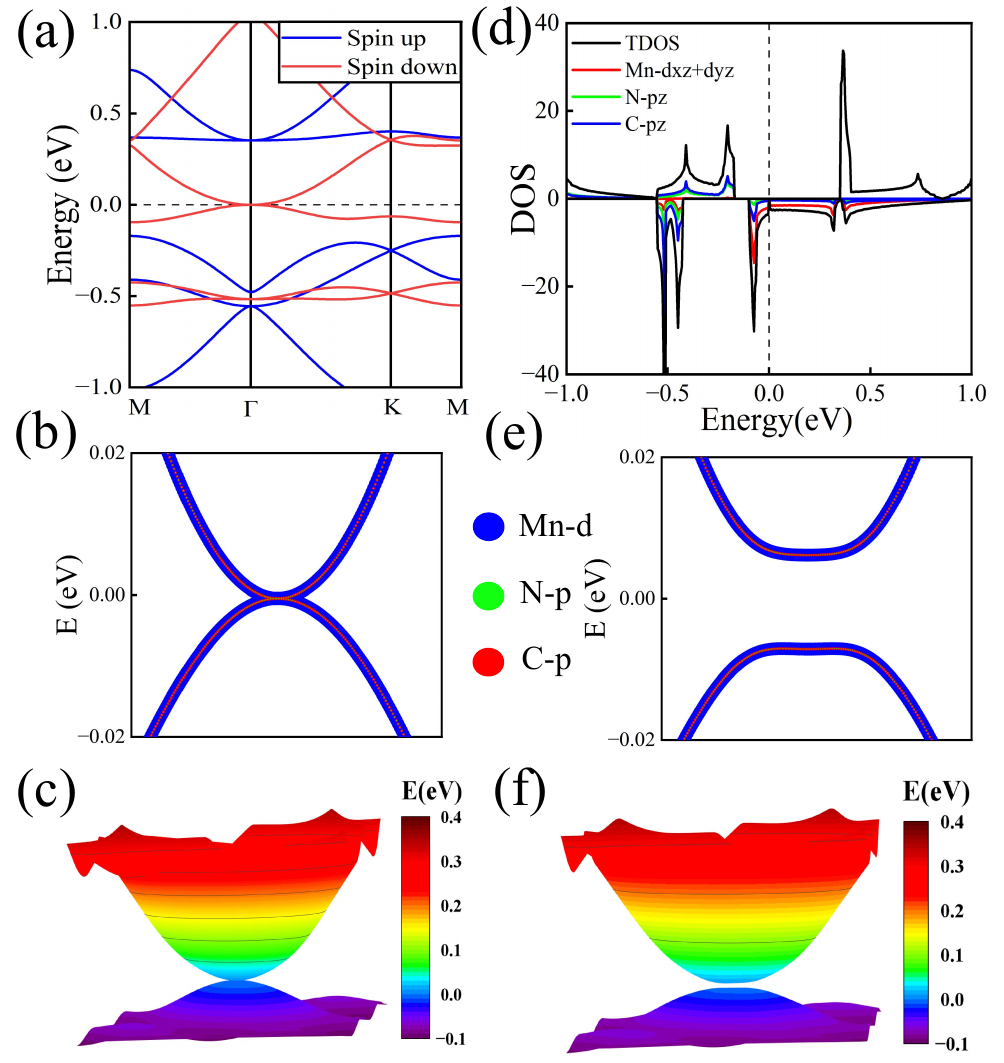}
\caption{\label{band_Mn} (a) The energy bands of monolayer (MnN$_4$)$_3$C$_{32}$ along the high-symmetry lines without SOC in the FM state. The spin-up and spin-down bands are colored in blue and red, respectively, and the Fermi level is set to zero. (b) and (e) give the zoomed-in view of the areas around $\Gamma$ point without and with SOC, respectively. (c) and (f) show the corresponding three-dimensional energy dispersion of bands near the Fermi level around $\Gamma$ point without and with SOC, respectively. (d) The TDOS and PDOS of monolayer (MnN$_4$)$_3$C$_{32}$. }
\end{figure}

Replacing the element Pt with the magnetic element Mn, the geometric structure of monolayer (MnN$_4$)$_3$C$_{32}$  is almost the same as that of monolayer (PtN$_4$)$_3$C$_{32}$, except for the optimized lattice parameters $a$ = $b$ = 12.32 $\AA$. Different with a nonmagnetic ground state of monolayer (PtN$_4$)$_3$C$_{32}$, monolayer (MnN$_4$)$_3$C$_{32}$ has a competing magnetic ground state involving multiple magnetic orders. Here we would like to focus on its ferromagnetic (FM) state. Figure \ref{band_Mn}(a) shows the calculated band structure of monolayer (MnN$_4$)$_3$C$_{32}$ in the absence of SOC in the FM state, in which the blue and red colors represent spin-majority and spin-minority channels, respectively. Ferromagnetic monolayer (MnN$_4$)$_3$C$_{32}$ is a perfect half-semi-metal, whose minority spin exhibits a quadratic bands-touching and vanishing density of states at the Fermi level while the majority one possesses a small, $\sim$0.52\ eV, insulating spin gap. A full (100\%) spin polarization near the Fermi level is thus expected. As the Fermi level lies almost in the middle of the spin gap, fully spin-polarized spin-filter efficiency can be maintained in a wide positive or negative bias range, which makes FM monolayer (MnN$_4$)$_3$C$_{32}$ an attractive candidate for spin-injection \cite{D0NA00530D,xx20}. Figure \ref{band_Mn}(b) illustrates the highest valence and lowest conduction bands of the minority spin touching each other at $\Gamma$ point at the Fermi level. These states are mainly composed of Mn-$d$, N-$p$, and C-$p$ orbitals. Meanwhile, the three-dimensional energy dispersion of these two bands is shown in Fig. \ref{band_Mn}(c).  A quadratic-crossing nodal cone is formed at $\Gamma$ point in the center of the first Brillouin zone. Similar to monolayer (PtN$_4$)$_3$C$_{32}$, there is a flat band that coexists with the nodal cone near the Fermi level in monolayer (MnN$_4$)$_3$C$_{32}$. The flat band locates right blow the Fermi level, which is fully occupied.

Figure \ref{band_Mn}(d) shows the density of states of monolayer (MnN$_4$)$_3$C$_{32}$ in the FM state. As expected, a vanishing density of states at the Fermi level is observed. These states near the Fermi level is composed of N-$p$, C-$p$, and Mn-$d$ orbitals. Moreover, from the PDOS,  it is clear that the C-$p_z$ and N-$p_z$ form a $\pi$ network together with the Mn $d_{xz}$ and $d_{yz}$ orbitals in the vicinity of the Fermi level. Since the flat band of (MnN$_4$)$_3$C$_{32}$ is quite close to Fermi level, correspondingly, there is a sharp peak shown in the TDOS, which is convenient for exotic quantum application like unconventional high temperature superconductivity.

Once SOC is taken into account, an energy gap of $\sim$13\ meV is opened at $\Gamma$ point, as shown in Fig. \ref{band_Mn}(e). Since Mn has a smaller atomic number than Pt, the effect of SOC in monolayer (MnN$_4$)$_3$C$_{32}$ is a little weaker in comparison with monolayer (PtN$_4$)$_3$C$_{32}$, although it is still much larger than that of graphene \cite{PhysRevB.75.041401}. Accompanying the opening of band gap, the quadratic-crossing nodal cone changes into a blunt cone with flat head, as shown in Figs. \ref{band_Mn}(e) and \ref{band_Mn}(f). The gap opening at $\Gamma$ point indicates the possible existence of topological nontrivial features to host a significant quantum anomalous Hall effect.

\begin{figure} 
\centering
\includegraphics[width=1.0\linewidth]{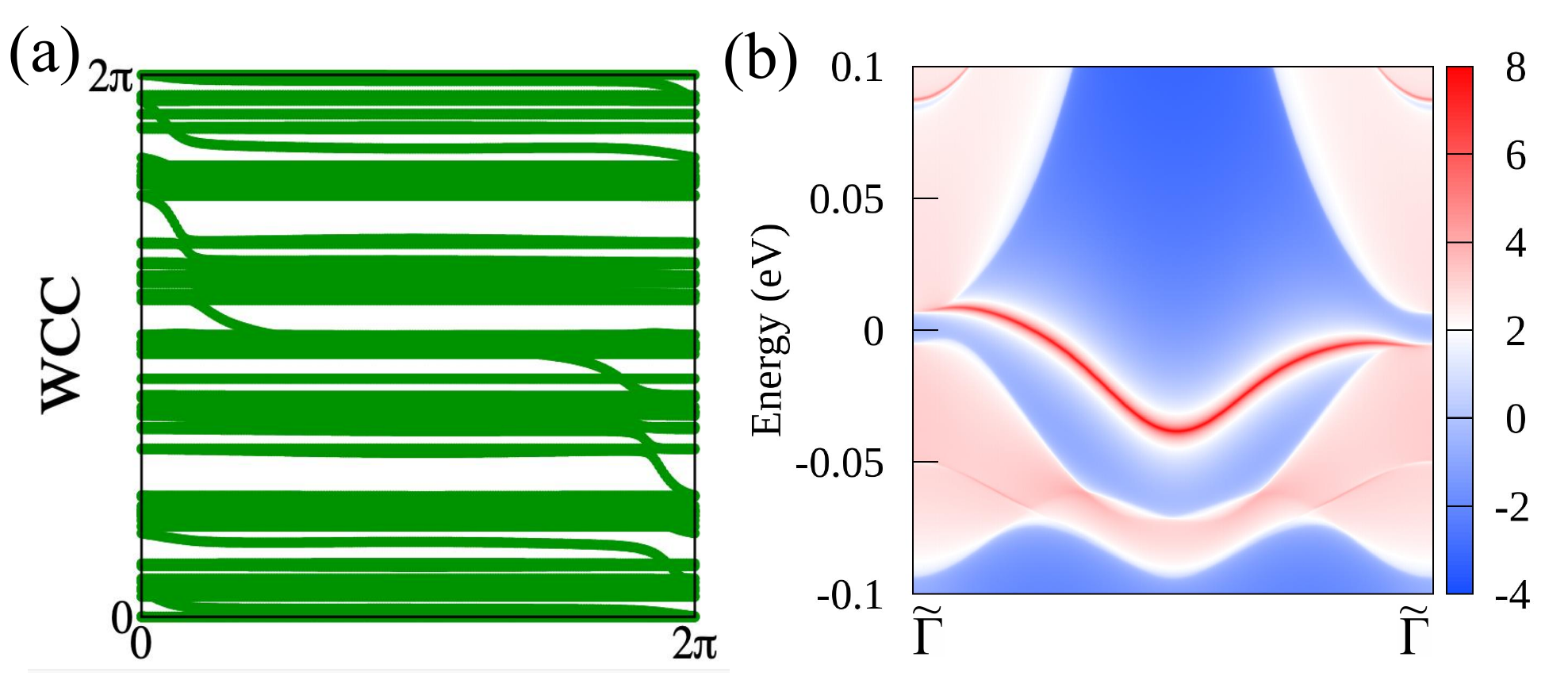}
\caption{\label{W90_Mn}  (a) The evolution of Wannier charge center (WCC) curves in Brillouin zone to calculate the Chern number invariant. (b) The chiral edge and bulk band structure of (MnN$_4$)$_3$C$_{32}$ projected along the (010) direction.}
\end{figure}

From the DFT band structure with SOC, we extract a tight-bind Hamiltonian of monolayer (MnN$_4$)$_3$C$_{32}$ in the basis of maximally localized wannier functions \cite{RevModPhys.84.1419,mostofi2008wannier90}. The band structure constructed by the means of maximally localized wannier functions agrees well with the DFT calculations, during which Mn-$d_{xz}$, Mn-$d_{yz}$, N-$p_z$, and C-$p_z$ orbitals are adopted as the projections. To further confirm the topological non-triviality of monolayer (MnN$_4$)$_3$C$_{32}$, we have calculated its  topological invariant Chern number. As shown in Fig. \ref{W90_Mn}(a), it is clear from the figure that the Wannier charge center evolution curves cut an arbitrary horizontal reference line odd times, indicating a value of $C =1$. Thus the gapped monolayer (MnN$_4$)$_3$C$_{32}$ with SOC is a two-dimensional Chern topological insulator.

We further investigate the chiral edge states of monolayer (MnN$_4$)$_3$C$_{32}$ projected along the (010) direction, whose band dispersions are given in Fig. \ref{W90_Mn}(b). One topological nontrivial edge state is visible,  which crosses the bulk band gap and connects the bulk conduction and valence bands of monolayer (MnN$_4$)$_3$C$_{32}$ at the two projected nodal-cone point in the Brillouin zone along the high-symmetry line $\widetilde{\Gamma}$--$\widetilde{M}$--$\widetilde{\Gamma}$. The results suggest that monolayer (MnN$_4$)$_3$C$_{32}$ is an ideal platform for exploring fascinating physical phenomena associated with novel 2D half-semi-metal and quantum anomalous Hall insulator, and realizing realistic spintronic devices. Meanwhile, it is also an ideal platform for exploring various topological phases of matter as low dimensionality provides unprecedented opportunities to manipulate the quantum states in low-cost electronic nanodevices \cite{PhysRevLett.113.236802,PhysRevB.100.125408}.

\section{SUMMARY}

In summary, based on symmetry analysis and the first-principles electronic structure calculations, we predict that the two novel KIH materials,  (PtN$_4$)$_3$C$_{32}$ and (MnN$_4$)$_3$C$_{32}$, are promising topological materials with fascinating electronic and topological properties. The coexistence of Dirac/quadratic-crossing cone and flat bands near the Fermi level in the KIH materials make them ideal for the high-tech applications in future. For (PtN$_4$)$_3$C$_{32}$,  it is a Dirac semimetal in the absence of SOC, while the system is converted into a $\mathbb{Z}_2$ topological quantum spin Hall insulator defined on a curved Fermi level with SOC included. Unlike (PtN$_4$)$_3$C$_{32}$, FM (MnN$_4$)$_3$C$_{32}$ belongs to a half-semi-metal, with the coexistence of fully spin-polarization and quadratic-crossing nodal states. Once SOC is considered, a sizable topological nontrivial band gap opens, and it becomes a quantum anomalous Hall insulator with the nonzero Chern number and chiral edge states. Our findings not only predict a family of new 2D KIH Dirac materials, but also provide an experimentally feasible platform to explore the new emergent topological quantum Hall insulators and their fascinating fundamental physics.

\section{ACKNOWLEDGMENTS}

This work was financially supported by the National Natural Science Foundation of China under Grants Nos. 12074040, 11974207, 12274255, and 11974194. F. Ma was also supported by the BNU Tang Scholar. The computations were supported by the Center for Advanced Quantum Studies, Beijing Normal University.

\bibliography{reference}

\end{document}